\documentclass[aps,pra,showkeys,twocolumn,superscriptaddress]{revtex4-2}
\usepackage{array}
\usepackage{booktabs}
\usepackage{tabu}
\usepackage{dcolumn}
\usepackage{amsmath}
\usepackage{amsfonts}
\usepackage{float}
\usepackage{amssymb}
\usepackage{graphicx,color}
\usepackage[colorlinks={true}]{hyperref}
\hypersetup{colorlinks=true,linkcolor=red,citecolor=blue,urlcolor=blue}
\usepackage{graphicx}
\usepackage{subfigure}
\usepackage{graphicx}
\usepackage{dcolumn}
\usepackage{bm}
\usepackage{pstricks}
\usepackage{braket}
\usepackage{orcidlink}

\def\be{\begin{equation}}
  \def\ee{\end{equation}}
\def\bea{\begin{eqnarray}}
\def\eea{\end{eqnarray}}
\def\f{\frac}
\def\n{\nonumber}
\def\l{\label}
\def\p{\phi}
\def\o{\over}
\def\R{\rho}
\def\pa{\partial}
\def\om{\omega}
\def\na{\nabla}
\def\P{\Phi}
\begin{document}

\title{Quantum Speed Limits in Qubit Dynamics Driven by Bistable Random Telegraph Noise: From Markovian to Non-Markovian Regimes}

\author{Maryam Hadipour \orcidlink{0000-0002-6573-9960}}
\email{maryam.hadipour@sci.uut.ac.ir}
\affiliation{Faculty of Physics, Urmia University of Technology, Urmia, Iran}

\date{\today}
\def\be{\begin{equation}}
  \def\ee{\end{equation}}
\def\bea{\begin{eqnarray}}
\def\eea{\end{eqnarray}}
\def\f{\frac}
\def\n{\nonumber}
\def\l{\label}
\def\p{\phi}
\def\o{\over}
\def\R{\rho}
\def\pa{\partial}
\def\om{\omega}
\def\na{\nabla}
\def\P{$\Phi$}

\begin{abstract}
We investigate the quantum speed limit (QSL) of a single superconducting qubit subjected to pure dephasing induced by bistable random telegraph noise (RTN), a common environmental disturbance in solid-state quantum systems. Using an exactly solvable model, we explore how the interplay between RTN parameters—the switching rate, coupling strength, and initial condition governs the transition between Markovian and non-Markovian dynamics. A coherence-based measure is employed to quantify non-Markovianity, and a unified quantum speed limit bound is derived based on relative purity. Our results reveal that in thermodynamic equilibrium, non-Markovian memory effects significantly reduce the quantum speed limit time, accelerating quantum evolution through information backflow. In contrast, under non-equilibrium initial conditions, the system exhibits purely Markovian behavior regardless of coupling strength, although strong coupling still leads to speedup via enhanced dephasing. These findings offer valuable insights for designing fast and noise-resilient quantum protocols by controlling both dynamical parameters and noise initialization.

\end{abstract}


\maketitle

\section{INTRODUCTION}	
The fundamental principles of the natural world impose an immutable and unavoidable constraint on the maximum velocity at which physical processes can occur, delineating a universal boundary inherent to the fabric of reality.  In quantum systems, the uncertainty principle establishes the quantum speed limit (QSL), constraining the velocity of processes at the subatomic scale. In quantum mechanics, the quantum speed limit time  is a measure that determines the maximum speed of a quantum system's dynamic evolution; it represents the shortest time required for a quantum system to transition from one state to another that is quantum-mechanically distinguishable. In closed systems, the minimum time required for a quantum state to evolve into an orthogonal state, known as the quantum speed limit, is determined by the energy variance or the average energy. This quantity is expressed in the Mandelstam-Tamm (MT) relation as $\tau_{qsl}= \pi \hbar/2 \Delta E$ and in the Margolus-Levitin (ML) relation as $\tau_{qsl}=\pi \hbar /2 \langle E \rangle$.

In realistic conditions, it is impossible to prevent the interaction between a quantum system and its surrounding environment; therefore, such a system must be modeled as an open quantum system \cite{i1}. In recent years, the investigation of the quantum speed limit in open quantum systems has emerged as a highly popular and significant topic of interest among researchers.
\cite{i2,i3,i4,i5,i6,i7,i8,i9,i10,i11,i12,i13,i14,i15,i16,i17,i18,i19,i20,i21,i22,i23,i24,i25,i26,i27,i28,i29,i30,kl} 
In their study \cite{i20}, Taddei et al. analyzed the quantum speed limit for open quantum systems by employing the quantum Fisher information bound, which pertains to the composite Hilbert space of the system and its environment. This approach builds upon a methodology previously developed in Ref. \cite{i21}. For open quantum systems exhibiting Lindblad-type evolution, Del Campo et. al utilized the concept of relative purity to derive a quantum speed limit equivalent to the Mandelstam-Tamm (MT) bound \cite{i3}. 
Deffner and Lutz, under the condition that the initial state of the quantum system is pure, employed the Bures angle to derive a unified bound for the quantum speed limit that encompasses both the Margolus-Levitin (ML) and Mandelstam-Tamm (MT) bounds in open systems. They further demonstrated that the presence of non-Markovian effects can accelerate the quantum evolution process \cite{i17}.

In addition to the aforementioned studies, extensive research has explored various aspects related to the quantum speed limit in open quantum systems, encompassing the dependence of the quantum speed limit on the initial state of the system \cite{i22}, the investigation of different types of quantum acceleration in longitudinal and transverse directions \cite{i23}, the analysis of decay rates in the deformed boson model with consideration of spin \cite{i24}, the impact of classical driving on enhancing the speed of quantum evolution \cite{i25}, the role of excited state populations \cite{i19}, the utilization of experimentally measurable criteria to determine the quantum speed limit \cite{i26}, the employment of alternative fidelity measures \cite{i27}, the use of relative purity as a criterion \cite{i28}, the consideration of conditions without relying on the rotating wave approximation \cite{i29}, the examination of the quantum speed limit in non-equilibrium environments \cite{i30}, and the analysis of the quantum speed limit in multipartite open systems \cite{i5}.

Environmental noise, such as random telegraph noise (RTN) induced by bistable impurities, significantly impacts the coherence and evolution of these systems \cite{i1}. Understanding the fundamental limits on how fast quantum states can evolve in the presence of such noise is crucial for optimizing quantum information processing tasks. In this work, we leverage an exactly solvable model of a single superconducting qubit subjected to pure-dephasing RTN, as introduced by Lo Franco et al. \cite{i31}. This model captures the crossover between Markovian and non-Markovian dynamics through the parameters $\lambda$ and $\gamma$, where $\lambda$ is the qubit-RTN coupling strength and $\gamma$ is the impurity switching rate. By analyzing the qubit’s coherence evolution under this noise model, we investigate the QSL time for a single qubit, exploring how the interplay between Markovian ($g<1$)and non-Markovian ($g>1$) regimes influences the fundamental bounds on quantum evolution speed. Our study aims to provide insights into the impact of realistic noise sources on the performance of superconducting qubits and to guide the design of noise-resilient quantum protocols.  It has been shown in \cite{i17} that the evolution of a quantum system can be accelerated by non-Markovian effects. It will also be showed that in considered model, the non-Markovianity  can accelerate the dynamic evolution of quantum systems. This effect arises due to the presence of memory in the environment and the feedback of quantum information from the environment to the system, which occurs in the non-Markovian regime. Unlike Markovian dynamics, where quantum information is unidirectionally lost to the environment, in the non-Markovian regime, reciprocal interactions between the system and the environment can lead to the preservation or even recovery of coherence, thereby accelerating the evolution of quantum states.
\section{The model}
The model involves single-qubit system which affected by a bistable impurity that generates random telegraph noise (RTN) at the pure dephasing operating condition. The total Hamiltonian of the system is is given by
\begin{equation}\label{H1}
H(t)=-\frac{1}{2}(\Omega+ \xi(t)) \sigma_z,
\end{equation}
where $\Omega$ is the intrinsic frequency of the qubit, $\sigma_z$ is the Pauli operator along the z-axis, $\xi(t)$ is the stochastic process generating random telegraph noise (RTN), switching between $\pm 1$ at rate $\gamma$ and $\lambda$ is the coupling strength between the qubit the qubit and RTN. This  stochastic process has the following statistical properties : $\langle \xi (t) \rangle = 0$ and $\langle \xi(t)\xi(t^\prime) \rangle = e^{-2 \gamma \vert t-t^{\prime} \vert}$. The spectral density $J(\omega)$ corresponding to the equilibrium fluctuations of $\xi(t)$ in the absence of external perturbations can be obtained through the Fourier transform of its autocorrelation function as 
\begin{equation}
J(\omega)=\int_{-\infty}^{\infty} \langle \xi(t) \xi(0)\rangle e^{i \omega t} dt = \frac{\lambda^2 \gamma}{2 (\gamma^2 + \omega^2)}.
\end{equation}
The key parameter in this model is the ratio $\lambda /\gamma$, which determines the transition between the Markovian $\lambda < \gamma $ and non-Markovian $\lambda > \gamma$ regimes. Since the Hamiltonian only involves $\sigma_z$, the diagonal elements of the density matrix (populations) remain constant, and only the off-diagonal elements (coherence) evolve.

To provide a geometric interpretation of the qubit dynamics under random telegraph noise (RTN) in the pure dephasing regime, we represent the qubit's initial state using the Bloch vector $\vec{r} = (r_x, r_y, r_z)$, where $r_x = \langle \sigma_x \rangle$, $r_y = \langle \sigma_y \rangle$, and $r_z = \langle \sigma_z \rangle$ are the expectation values of the Pauli operators. The initial density matrix is expressed as:

\begin{equation}
\rho(0) = \frac{1}{2} \left( I + \vec{r} \cdot \vec{\sigma} \right),
\end{equation}

with $\vec{\sigma} = (\sigma_x, \sigma_y, \sigma_z)$. The Bloch vector components satisfy $|\vec{r}| \leq 1$, with equality for pure states.

Given an initial Bloch vector $\vec{r}(0) = (r_x(0), r_y(0), r_z(0))$, the evolution of the Bloch vector components is described by

\begin{eqnarray}
r_x(t) &=& r_x(0) f(t) \cos \left( (\Omega + v/2) t \right)  \\
&+& r_y(0) f(t) \sin \left( (\Omega + v/2) t \right), \nonumber \\
r_y(t) &=& -r_x(0) f(t) \sin \left( (\Omega + v/2) t \right) \nonumber \\
&+& r_y(0) f(t) \cos \left( (\Omega + v/2) t \right), \nonumber \\
r_z(t) &=& r_z(0). \nonumber
\end{eqnarray}

where the decay factor $f(t)$ is:

\begin{equation}
f(t) = \left| A e^{-\frac{\gamma t (1 + \alpha)}{2}} + (1 - A) e^{-\frac{\gamma t (1 - \alpha)}{2}} \right|.
\label{eq:ft}
\end{equation}
where $A = \frac{1}{2\alpha} (1 + \alpha - i g \delta p_0)$, $\alpha = \sqrt{1 - g^2}$, and $\delta p_0$ represents the initial condition of the RTN ($\delta p_0 = 0$ for thermodynamic equilibrium or the values $\delta p_0 =\pm 1$).
\section{Quantum speed limit time}
The quantum speed limit (QSL) quantifies the minimal time required for a quantum system to evolve between two states, representing the maximal evolution speed. For open quantum systems with a mixed initial state, we adopt a distance measure based on the relative purity, as introduced by Campaioli et al. \cite{i32}. This measure defines the statistical distance between an initial state $\rho_0$ and an evolved state $\rho_\tau$ as 
\begin{equation}
\Theta\left(\rho_0, \rho_\tau\right)=\arccos \left(\sqrt{\frac{\operatorname{tr}\left[\rho_0 \rho_\tau\right]}{\operatorname{tr}\left[\rho_0^2\right]}}\right).
\end{equation}
This distance metric is particularly suitable for mixed states and provides a robust framework for deriving the QSL in nonunitary dynamics.

For an open quantum system governed by a time-dependent nonunitary evolution $\dot{\rho}_\tau = \mathcal{L}_\tau \rho_\tau$, QSL time is determined by a unified bound that generalizes the Mandelstam-Tamm (MT) and Margolus-Levitin (ML) types of bounds. The unified QSL bound is expressed as
\begin{equation}
\tau_{\mathrm{QSL}}=\max \left\{\frac{1}{\Lambda_\tau^{\mathrm{op}}}, \frac{1}{\Lambda_\tau^{\mathrm{tr}}}, \frac{1}{\Lambda_\tau^{\mathrm{bs}}}\right\} \sin ^2\left[\Theta\left(\rho_0, \rho_\tau\right)\right] \operatorname{tr}\left[\rho_0^2\right],
\end{equation}
where the terms $\Lambda_\tau^{op}$, $\Lambda_\tau^{tr}$ and $\Lambda_\tau^{hs}$ are time-averaged norms of the generator $\mathcal{L}_\tau(\rho_\tau)$,defined as
\begin{eqnarray}
\Lambda_\tau^{\mathrm{op}}&=&\frac{1}{\tau} \int_0^\tau\left\|L_\tau\left(\rho_\tau\right)\right\|_{\mathrm{op}} d t \\
\Lambda_\tau^{\mathrm{tr}}&=&\frac{1}{\tau} \int_0^\tau\left\|L_\tau\left(\rho_\tau\right)\right\|_{\mathrm{tr}} d t \nonumber \\
\Lambda_\tau^{\mathrm{hs}}&=&\frac{1}{\tau} \int_0^\tau\left\|L_\tau\left(\rho_\tau\right)\right\|_{\mathrm{hs}} d t . \nonumber
\end{eqnarray} 
Here, $\vert \vert . \vert \vert_{op}$, $\vert \vert . \vert \vert_{tr}$ and $\vert \vert . \vert \vert_{hs }$ denote the operator norm, trace norm, and Hilbert-Schmidt norm, respectively. The operator norm provides the tightest (ML-type) bound, as it satisfies the inequality $\Lambda_\tau^{op} \leq \Lambda_\tau^{hs} \leq \Lambda_\tau^{tr}$, making it the most effective measure for determining the QSL. This unified bound accounts for the initial state's mixedness through $tr[\rho_0^2]$ and is applicable to a wide range of open quantum systems, enabling the study of QSL in practical scenarios involving decoherence and environmental interactions.

Following the definition of the quantum speed limit (QSL) time, we introduce a measure of non-Markovianity based on quantum coherence, which is particularly relevant for open quantum systems. Quantum coherence, quantifying the superposition in the initial state, is defined using the $l_1$-norm as $\mathcal{C}(\rho)= \sum_{i\neq j} \vert \rho_{ij} \vert$, where $\rho_{ij}$ are the off-diagonal elements of the density matrix $\rho$ For a qubit in the Bloch representation, this simplifies to $\mathcal{C}\left(\rho_0\right)=\sqrt{r_x^2+r_y^2}$. Non-Markovianity, characterized by environmental memory effects, can be quantified by the non-monotonic evolution of quantum coherence under a quantum channel. In Markovian dynamics, coherence typically decreases monotonically due to decoherence. In contrast, non-Markovian dynamics may cause coherence to increase temporarily, reflecting information backflow from the environment. A measure of non-Markovianity is thus defined as the total increase in coherence over time, given by

\begin{equation}
\mathcal{N}_{\mathrm{coh}}=\int_{\dot{\mathcal{C}}(t)>0} \dot{\mathcal{C}}(t) d t,
\end{equation}
where $\dot{\mathcal{C}}(t)=\frac{d}{d t} \mathcal{C}\left(\rho_t\right)$ is the time derivative of coherence for the evolved state $\rho_t$, and the integral is taken over intervals where $\dot{\mathcal{C}}(t)>0$. A non-zero $\mathcal{N}_{\mathrm{coh}}$ indicates non-Markovian behavior, with larger values corresponding to stronger memory effects. In the context of the QSL, this coherence-based measure highlights how non-Markovianity influences evolution speed. For instance, in the damped Jaynes-Cummings model, strong non-Markovian effects (e.g., when the reservoir’s spectral width is small compared to the coupling strength) can enhance coherence temporarily, accelerating evolution, while high initial coherence tightens the QSL bound. Similarly, in the dephasing model, non-Markovianity drives coherence dynamics, scaling the QSL with $\mathcal{C}(\rho_0)$. This measure provides a practical tool to quantify non-Markovianity in systems where coherence is a critical resource, aiding the analysis of quantum protocols under environmental interactions.
\\
\begin{figure}[h]
\centering
    \includegraphics[width =0.7\linewidth]{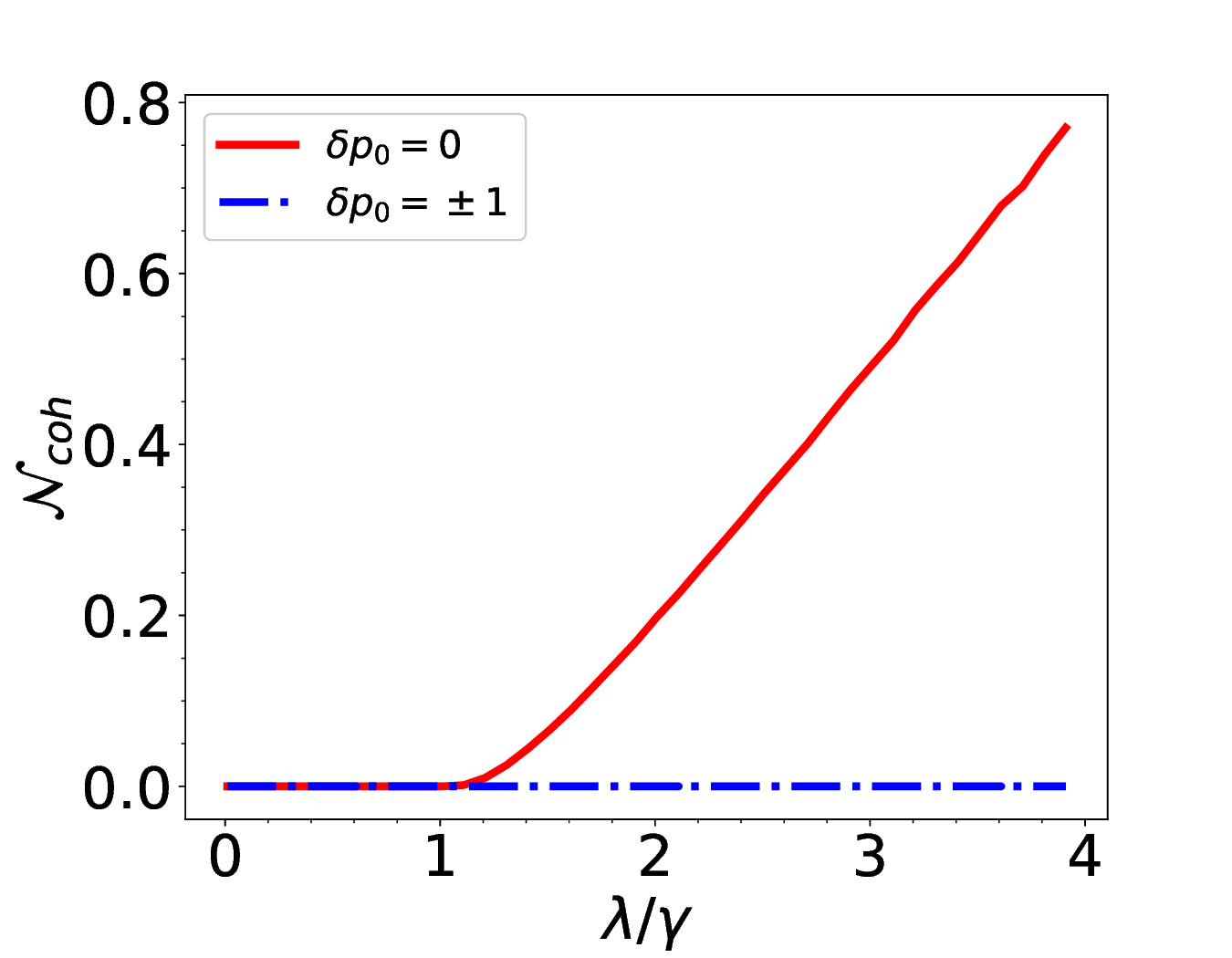}
    \vspace*{-5mm}
    \caption{Non-Markovianity $\mathcal{N}$ as functions of $\lambda/\gamma$. }
    \label{Fig1}
\end{figure}

Fig.\ref{Fig1} shows the Non-Markovianity  as a function of $\lambda/\gamma$ for two different types of the initial condition of the random telegraph noise, $\delta p_0$. The plot should show a sharp increase in non-Markovianity around $\lambda /\gamma=1$ for $\delta p_0=0$, marking the crossover from Markovian to non-Markovian dynamics. The non-Markovianity is equal to zero for $\delta p_0=\pm 1$, because these initial RTN states likely cause the coherence decay factor $f(t)$
 to decrease monotonically. This may result from phase effects in the complex parameter $A$, which suppress coherence revivals, or from dynamics that mimic Markovian behavior by stabilizing the noise’s dephasing effect. This result suggests that non-equilibrium RTN initial conditions ($\delta p_0 = \pm 1$) reduce environmental memory effects, potentially slowing quantum evolution and impacting qubit performance. Additional plot details or manuscript context would help confirm the exact mechanism.
\begin{figure}[h]
\centering
    \includegraphics[width =0.9\linewidth]{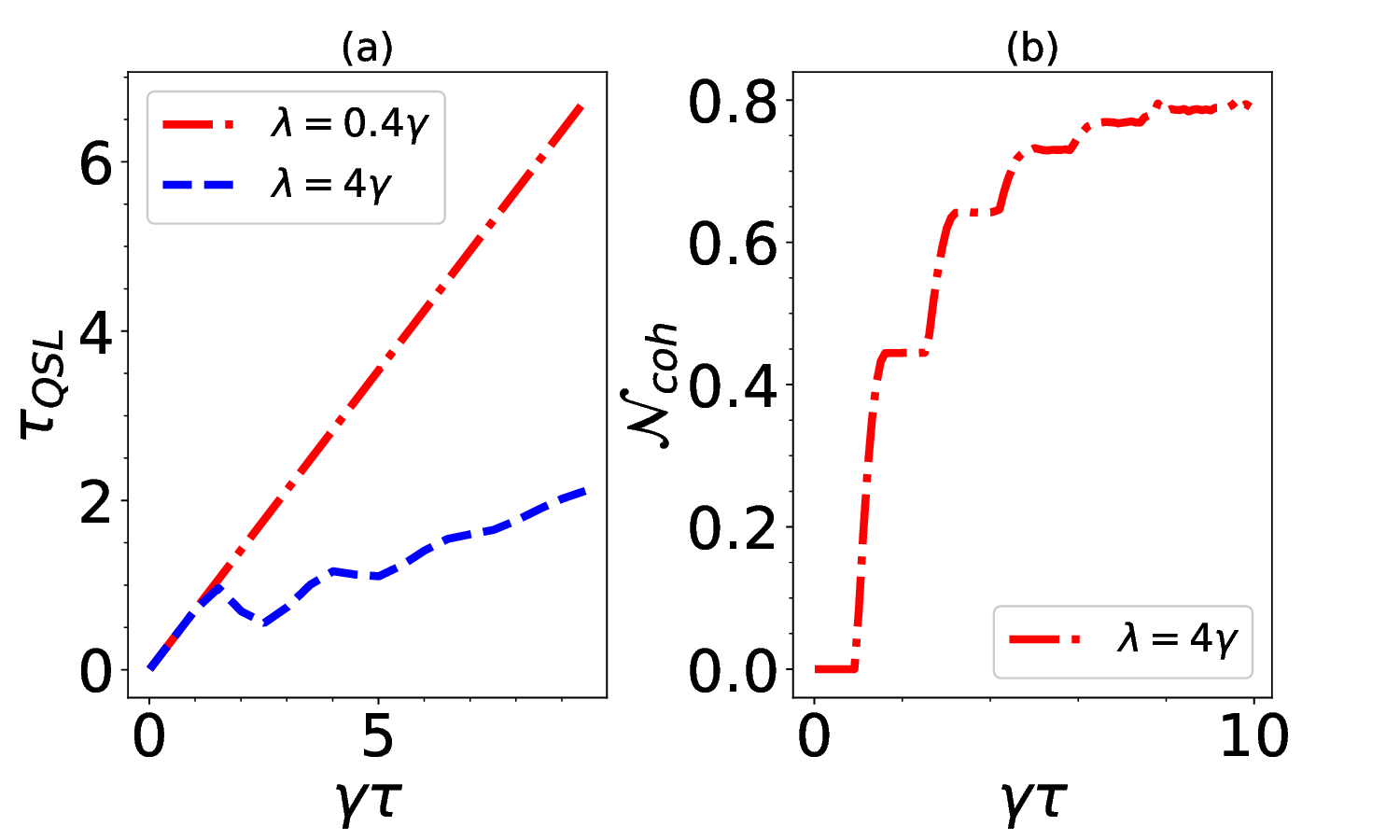}
    \vspace*{-5mm}
    \caption{(a) Quantum speed limit time in thermodynamic equilibrium ($\delta p_0=0$)in terms of dimensionless parameter $\gamma \tau$  for different value of $\lambda / \gamma$ with $r_x=r_y=r_z =0.5$. (b)Non-Markovianity $\mathcal{N}_{coh}$ in terms of dimensionless parameter $\gamma \tau$ in strong coupling regime.  }
    \label{Fig2}
\end{figure}
Fig. \ref{Fig2} illustrates the impact of environmental memory, in terms of Markovian and non-Markovian dynamics, on the quantum speed limit (QSL) time in thermodynamic equilibrium. The figure consists of two parts, which explore the system's behavior under different coupling regimes and dynamical conditions. In Fig. \ref{Fig2}(a), the quantum speed limit time $\tau_{QSL}$ is plotted as a function of the dimensionless parameter $\gamma \tau$ for various values of the ratio $\lambda / \gamma$. This ratio, representing the strength of coupling between the quantum system and its environment, is selected based on the analysis provided in Fig. \ref{Fig1}. In the non-Markovian regime $\lambda = 4 \gamma$ (with memory), the quantum speed limit time is significantly shorter than in the Markovian regime $\lambda=0.4 \gamma$ (without memory). This highlights the role of environmental memory in accelerating quantum dynamics. The reduction in QSL time in the non-Markovian regime may be attributed to the feedback of quantum information from the environment to the system, a hallmark of non-Markovian dynamics. Comparing Figs. \ref{Fig2}(a) and \ref{Fig2}(b)  provides further insights into the influence of coupling strength and environmental memory. Before the onset of non-Markovian behavior: In the early stages of evolution, the quantum speed limit time is identical for both strong coupling  and weak coupling  regimes. This indicates that in the Markovian regime or at short timescales, environmental memory effects have not yet become dominant. After the onset of non-Markovian behavior: As time progresses and environmental memory effects emerge, the QSL time in the strong coupling regime becomes significantly shorter than in the weak coupling regime and the memoryless case. This is due to stronger system-environment interactions in the non-Markovian regime, which lead to faster dynamics. The analysis of Figure 2 demonstrates that environmental memory, particularly in the non-Markovian regime and under strong coupling, significantly reduces the quantum speed limit time. This finding underscores the importance of non-Markovian dynamics in open quantum systems and suggests potential applications in quantum technologies, such as quantum computing \cite{i33} and quantum sensing \cite{i34,i35}, where rapid and efficient evolution is critical.
\begin{figure}[h]
\centering
    \includegraphics[width =0.7\linewidth]{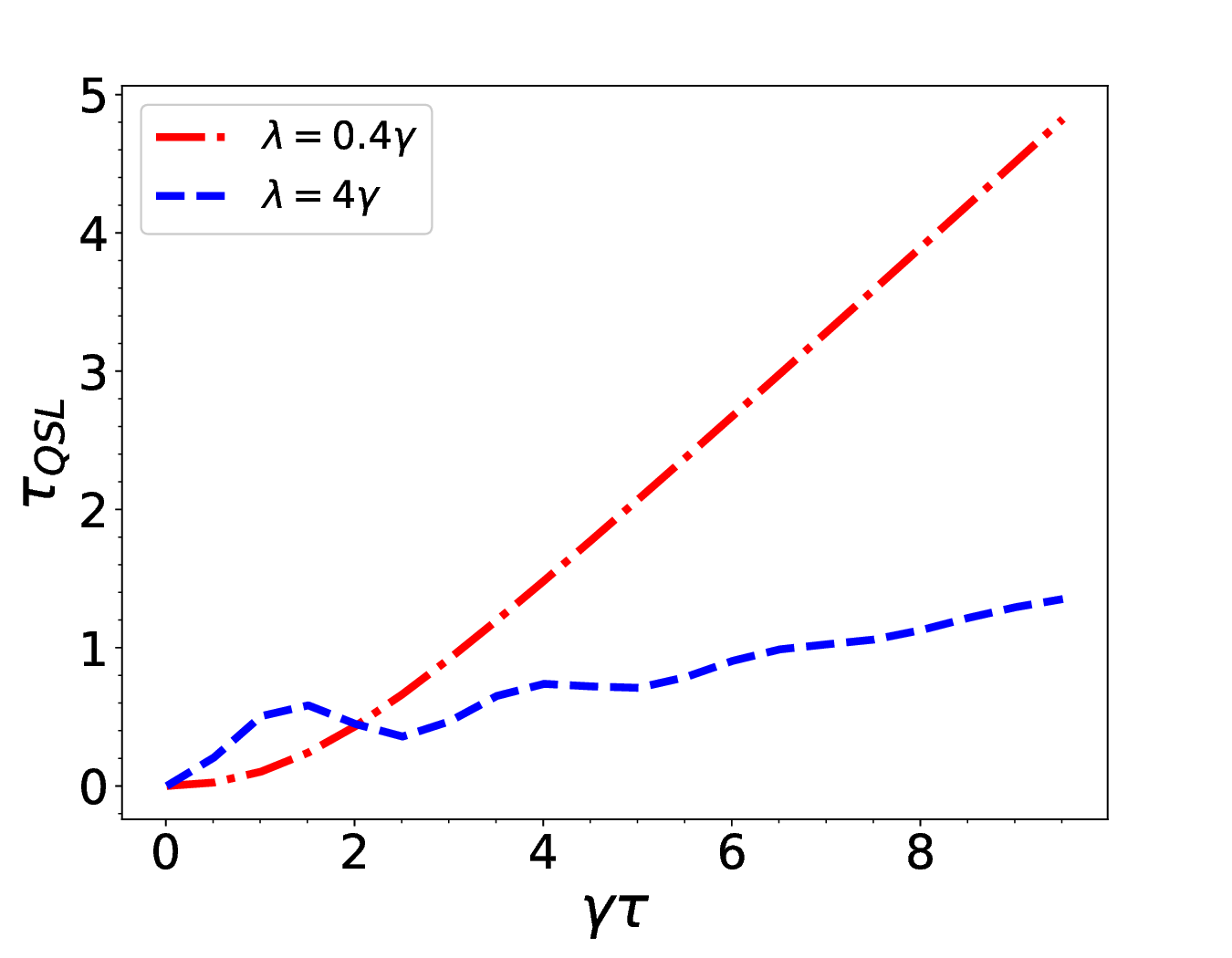}
    \vspace*{-5mm}
    \caption{(a) Quantum speed limit time in thermodynamic equilibrium ($\delta p_0=\ pm 1$)in terms of dimensionless parameter $\gamma \tau$  for different value of $\lambda / \gamma$ with $r_x=r_y=r_z=0.5$. (b)Non-Markovianity $\mathcal{N}_{coh}$ in terms of dimensionless parameter $\gamma \tau$ in strong coupling regime.  }
\label{Fig3}
\end{figure}
Fig. \ref{Fig3} illustrates the quantum speed limit (QSL) time $\tau_{QSL}$ under non-equilibrium initial conditions $\delta p_0 = \pm 1$ for various values of the coupling ratio $\lambda /\gamma$. According to Fig. \ref{Fig1}, under non-equilibrium initial conditions, the measure of non-Markovianity is zero for all values of $\lambda /\gamma$. This indicates that even in the strong coupling regime, the system’s dynamics remain entirely Markovian (memoryless).  However, increasing the coupling strength $\lambda /\gamma$ still reduces the quantum speed limit time compared to weak coupling. This reduction is due to a higher dephasing rate and stronger system-environment interactions, which accelerate quantum evolution. These results underscore the independent role of coupling strength as a factor in determining the speed of quantum evolution, beyond memory effects, and have potential applications in quantum technologies.
\begin{figure}[h]
\centering
    \includegraphics[width =0.9\linewidth]{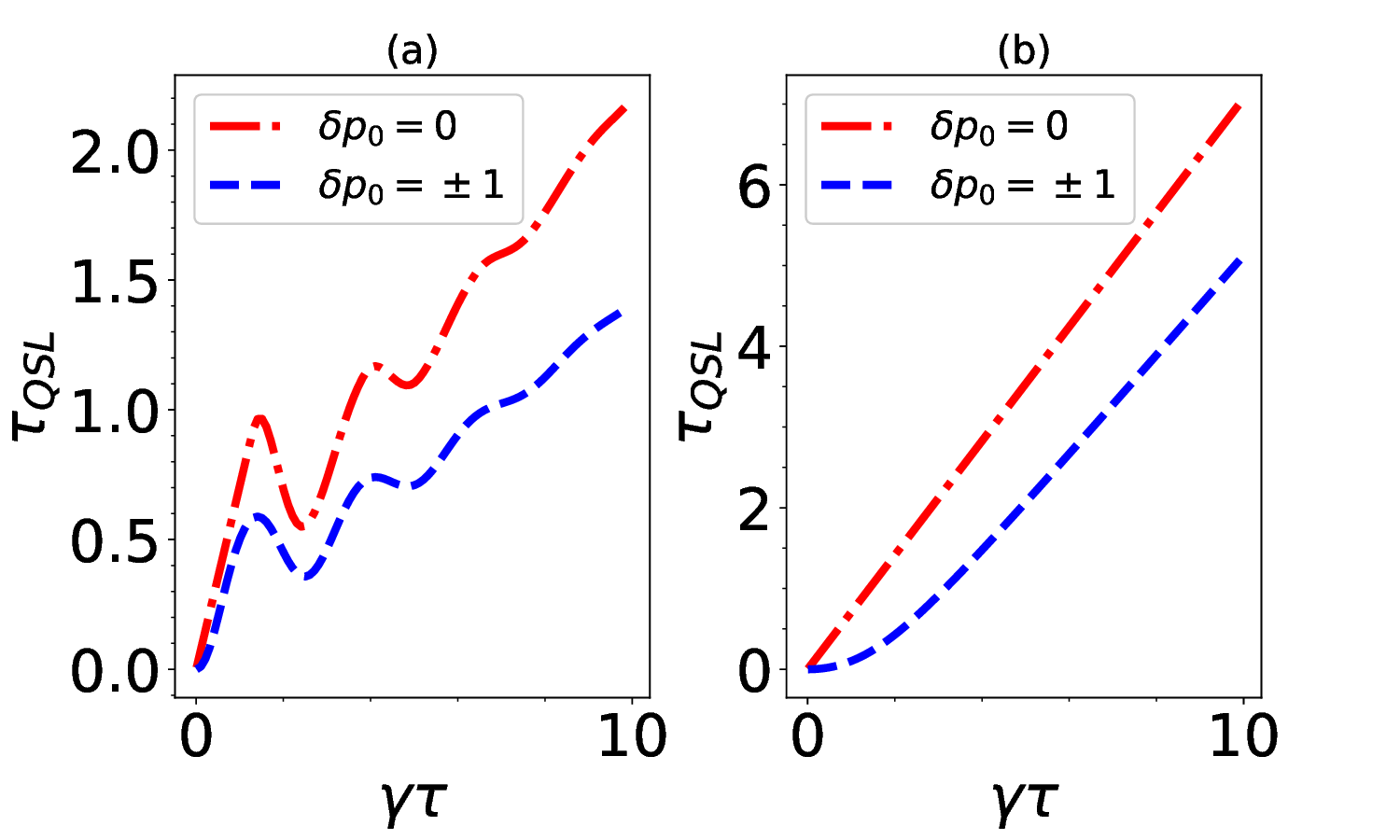}
    \vspace*{-5mm}
    \caption{(a) Quantum speed limit time in strong coupling regime $\lambda= 4 \gamma$ in terms of dimensionless parameter $\gamma \tau$  for different value of $\delta p_0$ with $r_x=r_y=r_z=0.5$. (b)Quantum speed limit time in weak coupling regime $\lambda= 0.4 \gamma$ in terms of dimensionless parameter $\gamma \tau$  for different value of $\delta p_0$ with $r_x=r_y=r_z=0.5$.}
\label{Fig4}
\end{figure}
Fig. \ref{Fig4} depicts the quantum speed limit  time  in both strong and weak coupling regimes for various values of the parameter $\delta p_0$. This figure examines the influence of initial conditions (equilibrium and non-equilibrium) on the speed of quantum evolution across different coupling regimes. In Fig. \ref{Fig4}(a), the quantum speed limit time is plotted as a function of the dimensionless parameter $\gamma t$ in the strong coupling regime $\lambda > \gamma$ for different values of $\delta p_0$. The parameter $\delta p_0$
determines the initial condition of the random telegraph noise (RTN), with $\delta p_0$ corresponding to thermodynamic equilibrium and $\delta p_0=\pm 1$ representing non-equilibrium conditions. When non-equilibrium initial conditions are present, the quantum speed limit time is significantly shorter than in thermodynamic equilibrium. This indicates that non-equilibrium initial conditions can accelerate quantum evolution in the strong coupling regime. This reduction in quantum speed limit time can be attributed to the influence of non-equilibrium initial conditions on dephasing dynamics.
Fig. \ref{Fig4}(b) shows that similar results are observed for the quantum speed limit (QSL) time in the weak coupling regime for different values of $\delta p_0$. It demonstrates that in the weak coupling regime, the quantum speed limit time under non-equilibrium initial condition  is shorter than in thermodynamic equilibrium , mirroring the results observed in the strong coupling regime. 
\begin{figure}[h]
\centering
    \includegraphics[width =0.7\linewidth]{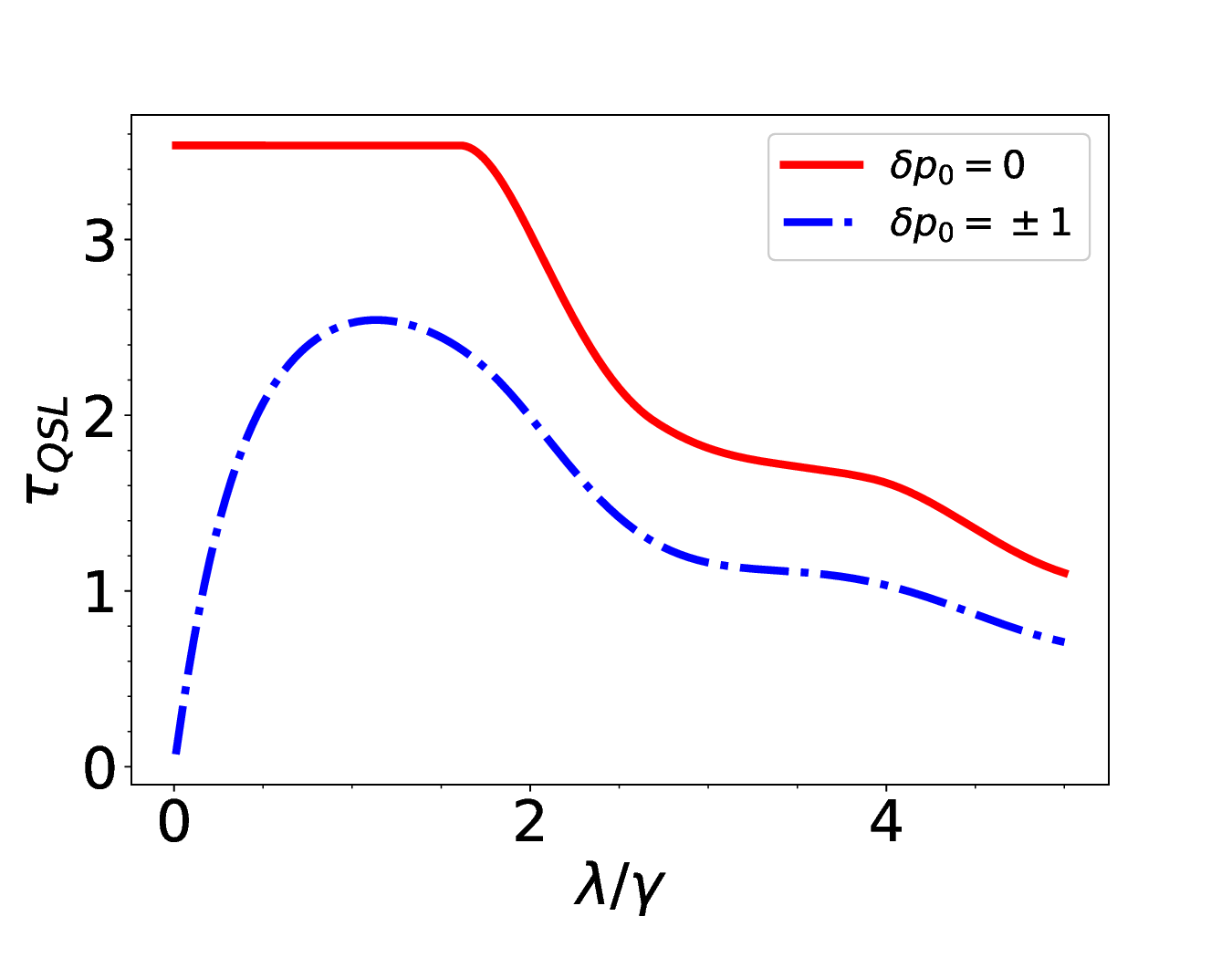}
    \vspace*{-5mm}
    \caption{(a) Quantum speed limit time in terms of coupling parameter $\lambda / \gamma$, with $r_x=r_y=r_z=0.5$ and $\gamma \tau =5$.  }
\label{Fig5}
\end{figure}
In Fig. \ref{Fig5}, the quantum speed limit time is plotted as a function of the ratio $\lambda / \gamma$, which indicates the strength of coupling, for two possible values of $\delta p_0$. It can be seen that, in the equilibrium case ($\delta p_0 =0$), increasing coupling and non-Markovian effects cause the system to evolve to the target state more quickly. This reduction in quantum speed limit time may result from an increased rate of information or energy exchange between the system and environment. Unlike the equilibrium state, dual behavior can be seen for quantum speed limit time. At weak coupling, the quantum speed limit time increases. This may occur because weak system-environment interactions limit the evolution rate, making quantum evolution slower. Dissipative effects gradually move the system away from its initial state, but this process is time-consuming due to the weak interaction. For larger $\lambda /\gamma$, as coupling increases, quantum speed limit time begins to decrease. In this regime, stronger interactions with the environment accelerate the system’s dynamics, likely due to increased dissipation or energy transfer rates. This behavior resembles the equilibrium state but lacks non-Markovian effects, so the quantum speed limit time reduction is solely due to stronger interactions.
\section{Results and discussion}
This study provides a comprehensive analysis of quantum speed limits in a qubit system influenced by random telegraph noise (RTN), focusing on how noise parameters—including the RTN's initial condition affect quantum evolution speed. Our results demonstrate that the choice of $\delta p _0$ has a profound effect on both the non-Markovian character of the dynamics and the quantum speed limit time. When the RTN is initialized in thermodynamic equilibrium $\delta p_0 =0$, non-Markovian behavior emerges for strong coupling $\lambda > \gamma$, leading to a significant reduction in QSL time due to environmental memory and information backflow. Conversely, for non-equilibrium initial conditions $\delta p_0 = \pm 1$, the evolution remains entirely Markovian even under strong coupling, and non-Markovianity vanishes. Despite this, increased coupling still reduces the QSL time due to enhanced dephasing rates, highlighting an independent role of coupling strength in accelerating dynamics. This dual dependence—on both memory effects and initial environmental states—underscores the importance of carefully preparing the noise environment in practical quantum technologies. Overall, our work suggests new routes to speed up quantum protocols by engineering initial conditions and system-environment interactions, with implications for quantum computing, metrology, and noise-resilient qubit design.

\section*{Acknowledgments}

\section*{Disclosures}
The authors declare that they have no known competing financial interests.

\section*{Data availability}
No datasets were generated or analyzed during the current study.



\end{document}